\documentstyle[ltwol,epsf]{article}

\def\be{\begin{equation}}
\def\ee{\end{equation}}
\def\bea{\begin{eqnarray}}
\def\eea{\end{eqnarray}}

\begin{document}

\title{$B\to K(K^*)X$: DECAY DISTRIBUTIONS, BRANCHING RATIOS AND CP 
ASYMMETRIES}

\author{CHANGHAO JIN}

\address{School of Physics, University of Melbourne, Victoria 3010,
Australia}   




\twocolumn[\maketitle\abstracts{ We discuss an improved theoretical description
of the semi-inclusive $B$ meson decays $B\to K(K^*)X$. The
decay distributions are calculated. Their branching ratios are found
to be appreciable. The CP asymmetries in the neutral $B$ modes
$\bar{B^0}\to K^-(K^{*-})X$ are sizable. An observation of direct CP violation
and a measurement of $\gamma$ may come from these neutral $B$ modes.}]

\section{Introduction}
The semi-inclusive charmless hadronic decays $B\to K(K^*)X$, where $X$ is a 
hadronic recoil
system containing no charmed and strange particles, are useful for studying 
direct CP violation and determining the weak phase 
$\gamma$\,\, \cite{browder}. 
Direct CP violation can also be studied with exclusive or inclusive 
charmless hadronic
$B$ decays. Although theoretical uncertainties in inclusive decays 
may be small, it is experimentally hard to identify final states
inclusively, while exclusive decays have a clear experimental 
signature, but the theoretical calculation is not as clean. 
Semi-inclusive hadronic
decays, which lie somewhere between inclusive and exclusive hadronic decays,
have a small theoretical uncertainty and a clear experimental signature, 
thereby providing an interesting channel for studying direct CP violation. 
Here\footnote{Contribution to BCP4, Ise-Shima, Japan, February 19-23, 2001}
we report on a recent study \cite{bkx} of $B\to K(K^*)X$ decays.  

Charmless hadronic $B\to K(K^*)X$ decays involve two types of amplitudes: 
$b\to u$
tree amplitudes and $b\to s$ penguin amplitudes. Direct CP violation arises 
due to the interference between two or more participating amplitudes with
different weak and strong phases for a single decay mode.
The effective Hamiltonian for charmless hadronic B decays with
$\Delta S = 1$ is given by
\begin{eqnarray}
H_{eff} &=&{G_F\over \sqrt{2}}
\{ V_{ub} V^*_{us}(c_1 O_1 +c_2 O_2 + \sum_{n=3}^{11} c_n O_n)\nonumber\\
&&+  V_{cb} V^*_{cs} \sum_{n=3}^{11} c_n O_n \},
\label{hamiltonian}
\end{eqnarray}
where $O_n$ are local four-quark and magnetic moment operators and 
$c_n$ the corresponding Wilson 
coefficients. Both $c_n$ and the matrix elements of $O_n$ depend on the 
renormalization scheme and
scale. The effective theory based on the Wilson operator product
expansion provides a framework to separate the short- and long-distance 
strong interaction. The short-distance strong interaction effects above
the scale $\mu\sim m_b$ are
incorporated in the Wilson coefficients $c_n$. The long-distance strong
interaction effects below the scale $\mu$ are encoded in the hadronic matrix 
elements of the local operators $<XK(K^*)|O_n|B>$.   

The short-distance coefficients $c_n$ are well known. They have been 
calculated up to
the next-to-leading-order corrections \cite{coef}. 
Factorization has often been employed to calculate hadronic matrix elements. 
Recent theoretical work \cite{beneke,li,brodsky}
has justified factorization in the heavy quark limit in the case
that the ejected particle from the $B$ decay is a light meson or an onium.  
Perturbative QCD corrections to 
factorization can be computed in the heavy quark limit.
The separation of short and long distance QCD using the effective 
Hamiltonian
together with QCD factorization, light cone expansion and heavy quark
effective theory leads to an improved theoretical description of  
$B\to K(K^*)X$ decays.

\section{Initial Bound State Effects}
We choose to study $\bar{B^0}\to K^-(K^{*-})X$ and 
$B^-\to \bar{K^0}(\bar{K^{*0}})X$. The factorized matrix elements for these
processes do not involve the transition form factors, eliminating a potential
theoretical uncertainty. Using the effective Hamiltonian (\ref{hamiltonian})
and QCD factorization \cite{beneke},
the differential decay rate for $B\to KX$ in the $B$ rest frame is given by
\begin{eqnarray}
&&d\Gamma 
= {1\over 2m_B} {d^3 {\bf P}_K\over (2\pi)^3 2 E_K}
(-2f_K^2) \\
&&\times \left \{ |\alpha|^2 P_K^\mu P_K^\nu (g_{\mu \alpha}g_{\nu\beta}
+ g_{\mu\beta}g_{\nu\alpha} - g_{\mu\nu} g_{\alpha\beta})
+ |\beta|^2 g_{\alpha\beta}\right \}\nonumber\\
&&\times \int d^4y\,\, e^{iy\cdot P_K} [\partial^\alpha\Delta_{q'}(y)] <B|\bar b(0)
\gamma^\beta U(0,y) b(y)|B>, \nonumber
\label{rate}
\end{eqnarray}
where $q'=u$ and $d$ for $\bar{B^0}$ and $B^-$, respectively.
After factorization the long-distance QCD effects on $B\to KX$ decays are 
contained in
two matrix elements: $<K|\bar s\gamma^\mu (1-\gamma_5)q'|0>= if_KP_K^\mu$
defining the decay constant $f_K$ for a $K$ meson and
$<B|\bar b(0)\gamma^\beta U(0,y) b(y)|B>$ incorporating the effect of the
initial $b$ quark 
bound state in the $B$ meson. The $K$ meson decay constant is known
from experiment \cite{PDG}: $f_K= 159.8\pm 1.5$ MeV. 
We need to know another hadronic matrix element
$<B|\bar b(0)\gamma^\beta U(0,y) b(y)|B>$.

We use the light cone expansion method to handle the initial bound state
effect. Since the dominant contribution to the decay rate comes from the light
cone region, the light cone expansion can be used to systematically 
calculate the matrix element. In the leading twist approximation,
$<B|\bar b(0)\gamma^\beta U(0,y) b(y)|B>= 
2P_B^\beta\int d\xi\, e^{-i\xi y\cdot P_B}f(\xi)$, where $f(\xi)$ is the
$b$-quark distribution function \cite{jp,gamma}.
We then obtain the decay distribution as a function of $E_K$
\begin{eqnarray}
{d\Gamma(B\to K X)\over d E_K}
={f^2_K E_K\over 2\pi m_B} \left (4|\alpha|^2 E_K^2 + |\beta|^2\right ) f({2E_K\over m_B}).
\label{kx}
\end{eqnarray}
Carrying out similar calculations, we obtain the differential decay rate for the $B\to K^* X$
decay
\begin{eqnarray}
{d\Gamma(B\to K^* X)\over d E_{K^*}}
={f^2_{K^*} E_{K^*}^3\over 2\pi m_B} 4|\alpha_*|^2 f({2E_{K^*}\over m_B}).
\label{kstar}
\end{eqnarray}

The distribution of $E_{K^{(*)}}$ is a delta function with the peak at 
$E_{K^{(*)}}=m_b/2$ in the free quark decay $b\to K(K^*)q'$ in the $b$ rest
frame. Gluon bremsstrahlung 
and initial bound state effect smear the spectrum.
The leading effect of the initial $b$-quark bound state is described by 
the distribution 
function $f(\xi)$. It is important to notice that the same distribution
function also encodes the leading nonperturbative QCD effects in
$B\to X\gamma$\,\,  \cite{gamma} and 
inclusive semileptonic decays $B\to Xl\bar\nu$\,\,  \cite{jp}.
Universality of the distribution function enhances predictive power:
The distribution function can be measured in one process and then used to 
make predictions for
all other processes. 

Several important properties of the distribution function
are already known in QCD \cite{jp,gamma}. 
When integrating $\xi$ from 0 to 1, $\int^1_0d\xi f(\xi)$ must
give 1 due to current conservation.
If the decay can be considered to be a free $b$ quark decay, then the $b$ 
quark field is given by $b(y) = e^{-iy\cdot P_b} b(0)$, one obtains
\begin{eqnarray}
f(\xi) = \delta(\xi - {m_b\over m_B}).
\label{new}
\end{eqnarray}
We can also estimate the
mean $<\xi> = \int^1_0 d\xi \xi f(\xi)$ and
the variance $\sigma^2 = \int^1_0 d\xi \xi^2 f(\xi)- 
<\xi>^2$ using heavy quark effective theory \cite{hqet}. They are given by \cite{gamma,jp}
\begin{eqnarray}
&&<\xi> = {m_b\over m_B} [1+ {5\over 6m^2_b} (\mu^2_\pi-\mu^2_g)],\nonumber\\
&&\sigma^2 = {\mu^2_\pi\over 3m_B^2},
\end{eqnarray}
where
\begin{eqnarray}
\mu^2_g &=& {1\over 4m_B}
<B|\bar h g_s G_{\mu\nu} \sigma^{\mu\nu} h|B>,
\nonumber\\
\mu^2_\pi &=& - {1\over 2m_B}<B|\bar h (iD_T)^2 h|B>,
\end{eqnarray}
are two parameters of heavy quark effective theory.
The small value for $\sigma^2$ implies that the distribution function is sharply 
peaked around $m_b/m_B$.

The detailed form of the distribution function is not yet known. We use the following
general parametrization for the distribution function for our numerical
analysis: 
\begin{eqnarray}
f(\xi) = N {\xi (1-\xi)^c\over [(\xi -a)^2 +b^2]^d},
\label{dis}
\end{eqnarray}
where $N$ is a normalization constant which guarantees
$\int^1_0 d \xi f(\xi) = 1$. The four parameters $(a,b,c,d)$ respect all the 
known properties of the distribution function.

We have also calculated the initial bound state effect by directly using 
heavy quark effective theory. The $1/m_b$ expansion relates the matrix
element to the parameters of heavy quark effective theory \cite{ma}.
In terms of the parameters of heavy quark effective theory, the decay rates are given by
\begin{eqnarray}
\Gamma(B\to K X)
&=& {f^2_K \over 8\pi}m_b [|\alpha|^2m_b^2
(1+{7\over 6}{\mu^2_g\over m_b^2}
-{53\over 6} {\mu^2_\pi\over m_b^2}) \nonumber\\
&&+|\beta|^2 (1-{\mu_\pi^2\over 2 m_b^2} + {\mu^2_g \over 2 m_b^2})], \\
\Gamma(B\to K^* X)
&=& {f^2_{K^*} \over 8\pi}m_b |\alpha_*|^2m_b^2 \nonumber\\
&&\times (1+{7\over 6}{\mu^2_g\over m_b^2}
-{53\over 6} {\mu^2_\pi\over m_b^2}).
\label{heqet}
\end{eqnarray}
The free quark decay $b\to K(K^*)q'$ results are reproduced in the heavy
quark limit.

\section{Results and Discussions}
We have calculated the decay distributions, CP-averaged branching ratios,
and CP asymmetries in $B\to K(K^*)X$ decays. The phenomenological
treatment has been improved with better theoretical understanding.
There are still several sources of theoretical uncertainties, including
the input parameters, the meson light cone distribution amplitudes, 
the distribution function, the renormalization scale, and
the unknown power corrections. All the results obtained should be understood
as valid within the theoretical uncertainties.

\begin{figure}
\centerline{\epsfysize=7truecm \epsfbox{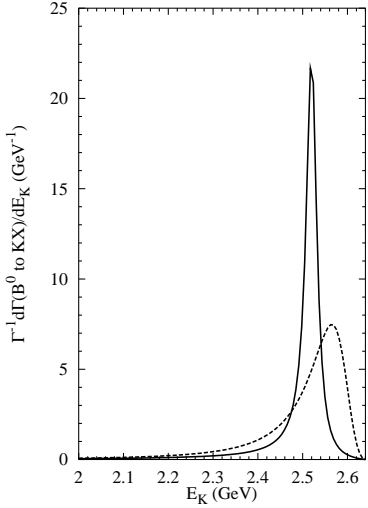}}
\caption{Kaon energy spectrum in $\bar{B^0}\to K^- X$. The solid and dashed curves 
correspond to two very different forms of the distribution function $f(\xi)$,
respectively.} 
\label{fig:spec}
\end{figure}

The distribution of the
$K(K^*)$ energy calculated in the heavy quark effective theory approach
is the same 
delta function with the peak at $E_{K^{(*)}}=m_b/2$ as in the free quark decay
$b\to K(K^*)q'$. The light cone expansion approach is capable of accounting for
the initial bound state effect on the decay distribution. The resulting
spectrum spreads over the full kinematic range 
$0\leq E_{K^{(*)}}\leq m_B/2$ and depends strongly on the form of the
distribution function. However, if the cut $E_{K^{(*)}}>2.1$ GeV is
applied to suppress the background, more than $97\%$ of events will
survive. For illustration, the kaon energy spectrum in $\bar{B^0}\to K^- X$ is  
shown in Fig.~1, computed in the light cone expansion approach
assuming $\gamma=60^\circ$.

One can also calculate the hadronic invariant mass spectrum 
$d\Gamma/dM_X$ in $B\to KX$. $M_X$ is related with $E_K$ in the $B$ rest
frame through $M_X^2= (P_B-P_K)^2= m_B^2-2m_BE_K+m_K^2$. The $M_X$ distribution
can be calculated by using $d\Gamma/dM_X= (M_X/m_B)d\Gamma/dE_K$ and the
above kinematic relation; $d\Gamma/dE_K$ is given in Eq.~(\ref{kx}).  
Similarly, one can also calculate the $M_X$ distribution in 
$B\to K^* X$.

We show the CP-averaged branching ratios, in Figs.~2-5, and the CP asymmetries, in Figs.~6-9,
in $B\to K(K^\ast) X$ as a function of the CP violating phase
$\gamma$. The solid curves are the results from the light cone
expansion, while
the dashed curves are from the free $b$ quark decay approximation.
The initial bound state effects
encoded in the distribution function cancel 
in the CP asymmetries in
$B\to K^\ast X$,
so that the solid and dashed curves coincide in Figs.~8 and 9.

We find that the
initial bound state effects on the branching ratios and
CP asymmetries are small.
In the light cone expansion approach, the CP-averaged branching ratios are
increased by about 2\% with respect to the free $b$-quark decay.
For $\bar B^0 \to K^- (K^{*-}) X$, the CP-averaged branching ratios
are sensitive to the phase $\gamma$
and the CP asymmetry can be as large as $7\%$ ($14\%$),
whereas for $B^-\to \bar K^0 (\bar K^{*0})X$ the CP-averaged branching ratios
are not sensitive to $\gamma$ and the CP asymmetries are small ($< 1\%$).
The CP-averaged branching ratios
are predicted to be in the ranges $(0.53 \sim 1.5)\times 10^{-4}$
[$(0.25 \sim 2.0)\times 10^{-4}$] for $\bar B^0 \to K^- (K^{*-})X$
and $(0.77 \sim 0.84)\times 10^{-4}$ [$(0.67 \sim 0.74)\times 10^{-4}$] for
$B^-\to \bar K^0 (\bar K^{*0}) X$, depending on the value of the CP violating
phase $\gamma$. In the heavy quark effective theory approach,
the branching ratios are decreased by about 10\% and the CP
asymmetries are not affected in comparison with the free $b$-quark decay.
The three estimates (free quark decay approximation, light cone expansion and
heavy quark effective theory method)
all give the same order of magnitudes for the branching
ratios and CP asymmetries. The branching ratios for $B\to K(K^*)X$ are of order
$10^{-4}$ and the CP asymmetries in the neutral $B$ modes $\bar{B^0}\to
K^-(K^{*-})X$ are sizable and can be measured at the $B$ factories.
\section*{Acknowledgments}
I would like to thank X.-G. He and J.P. Ma for their collaboration,
B. Stech for discussions,
and A.I. Sanda and his colleagues for their hospitality and
the excellent organization of the conference.
This work was supported by the Australian Research Council.

\begin{figure}
\centerline{\epsfysize=4truecm \epsfbox{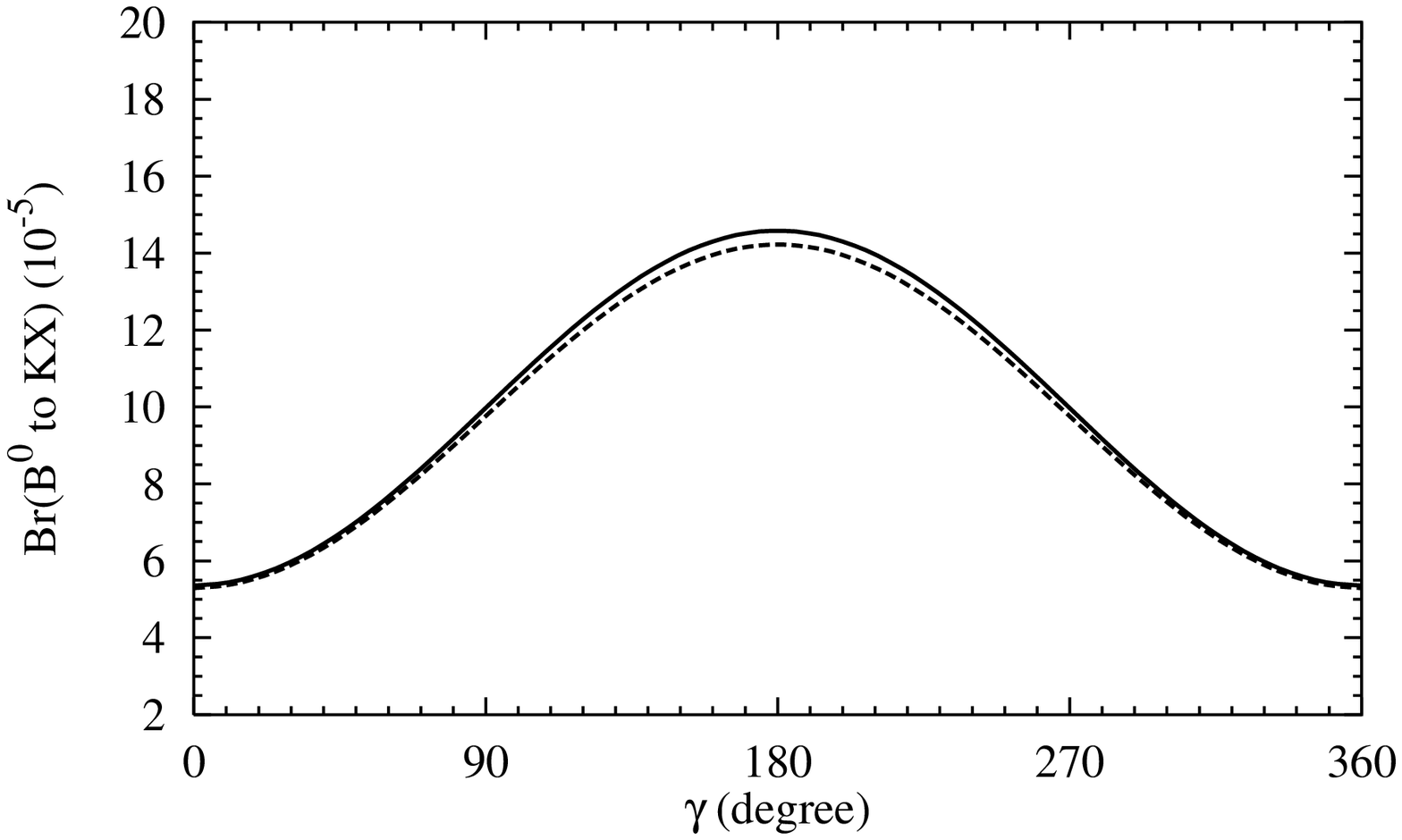}}
\caption {
CP-averaged branching ratio for $\bar B^0 \to K^- X$.}
\label{b0br} 
\end{figure}
\begin{figure}
\centerline{\epsfysize=4truecm \epsfbox{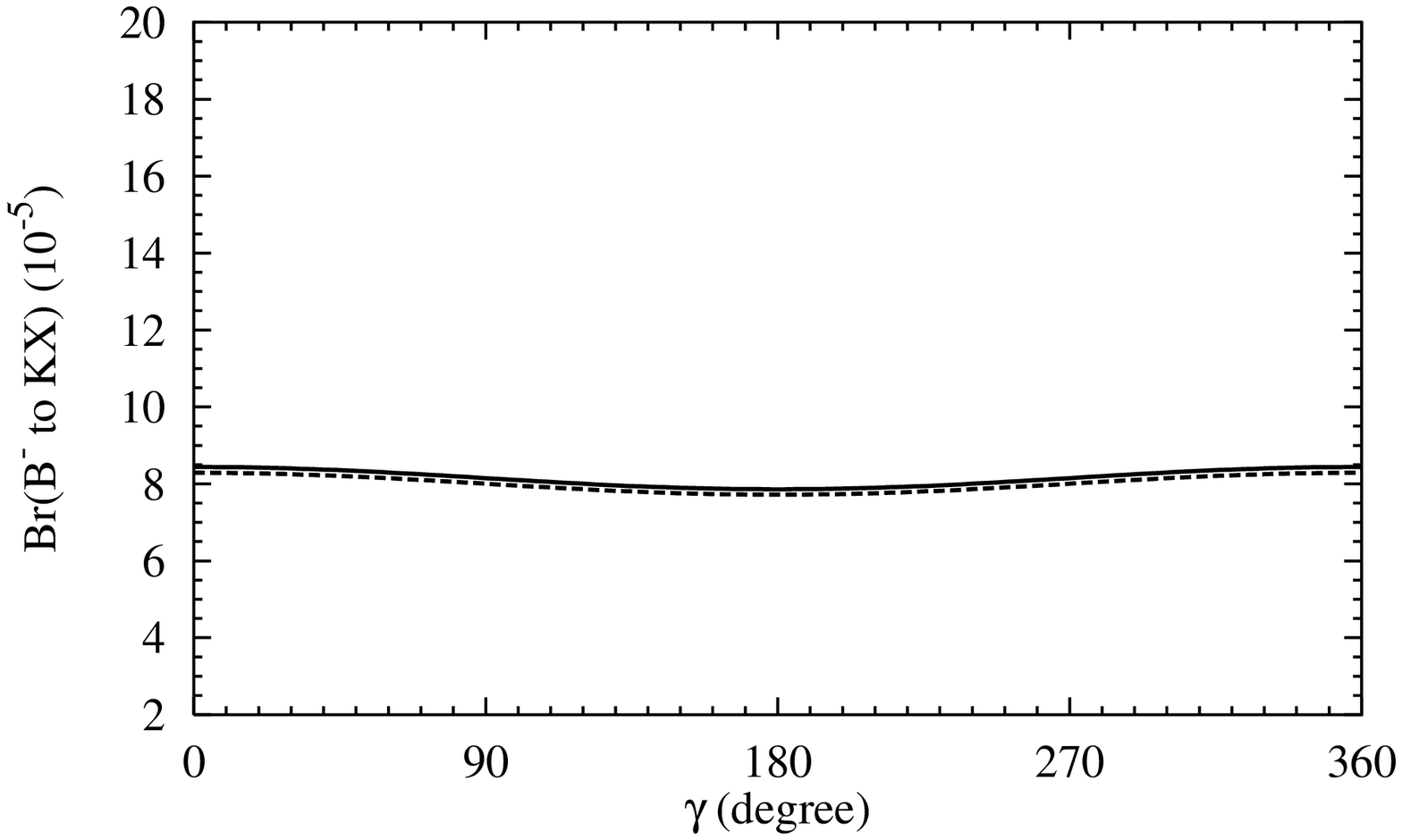}}
\caption {
CP-averaged branching ratio for $B^-\to \bar K^0 X$.
} \label{bmbr} 
\end{figure}
\begin{figure}
\centerline{\epsfysize=4truecm \epsfbox{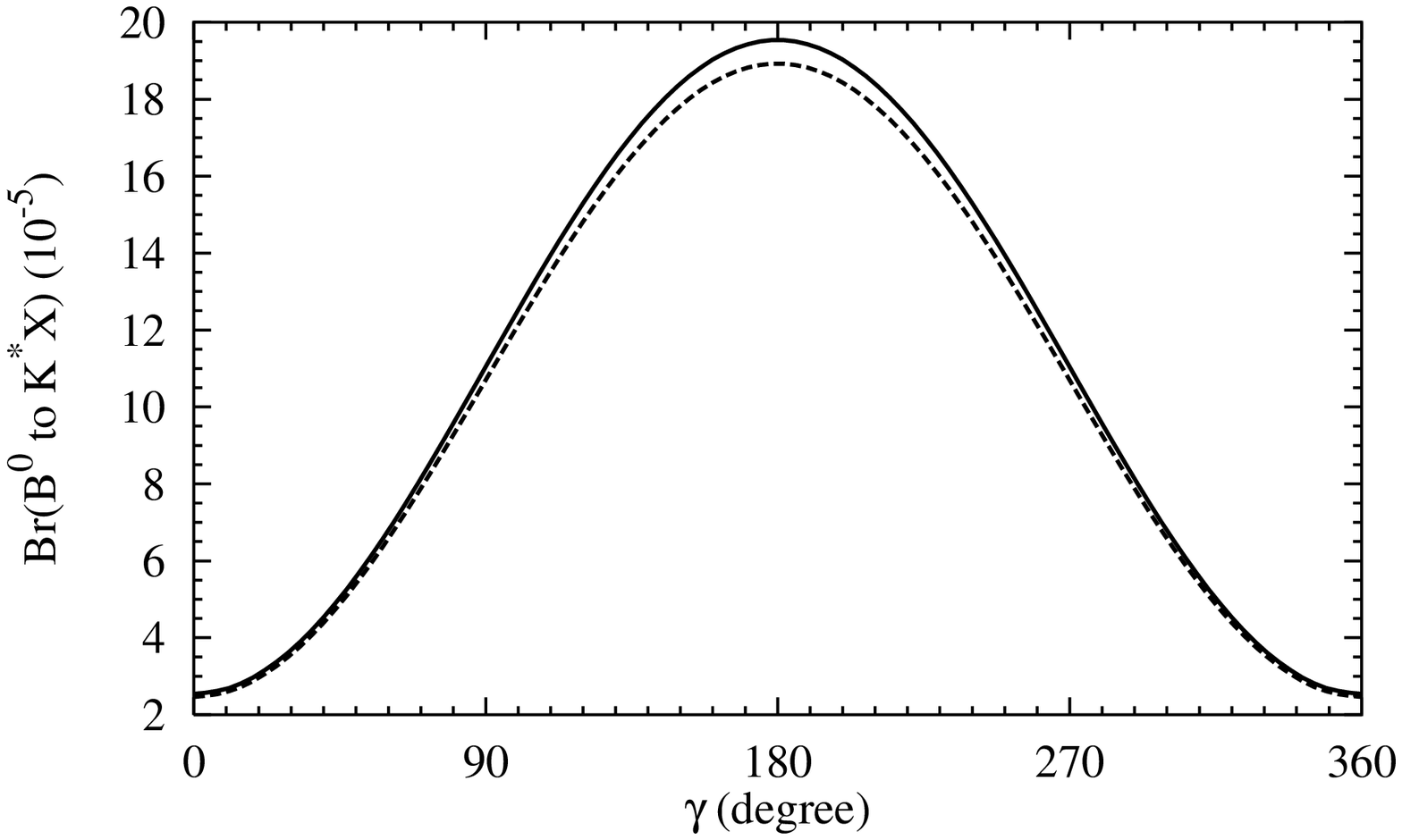}}
\caption {
CP-averaged branching ratio for $\bar B^0 \to K^{*-} X$.
} \label{b0starbr} 
\end{figure}
\begin{figure}
\centerline{\epsfysize=4truecm \epsfbox{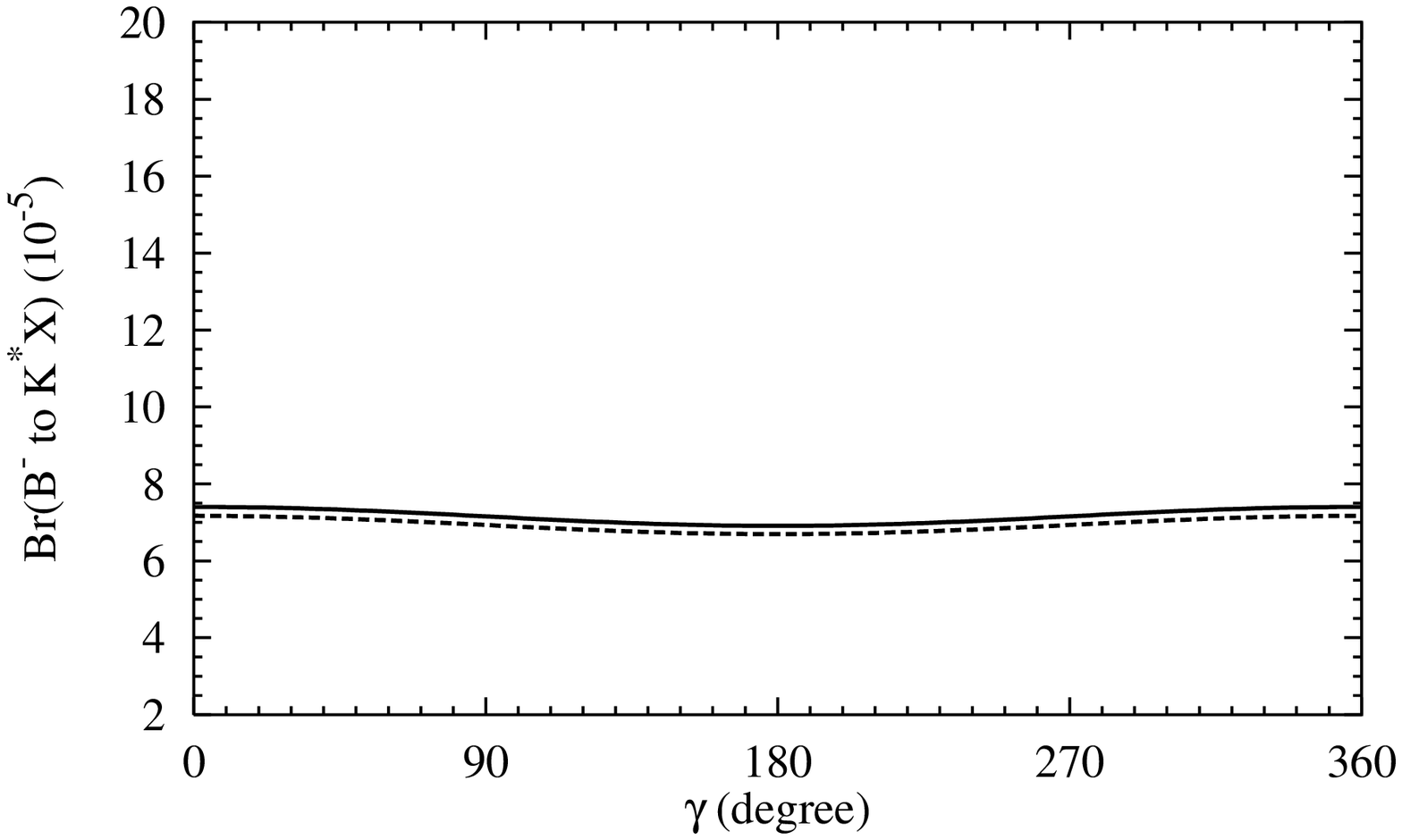}}
\caption {
CP-averaged branching ratio for $B^-\to \bar K^{*0} X$.
} \label{bmstarbr}
\end{figure}
\begin{figure}
\centerline{\epsfysize=4truecm \epsfbox{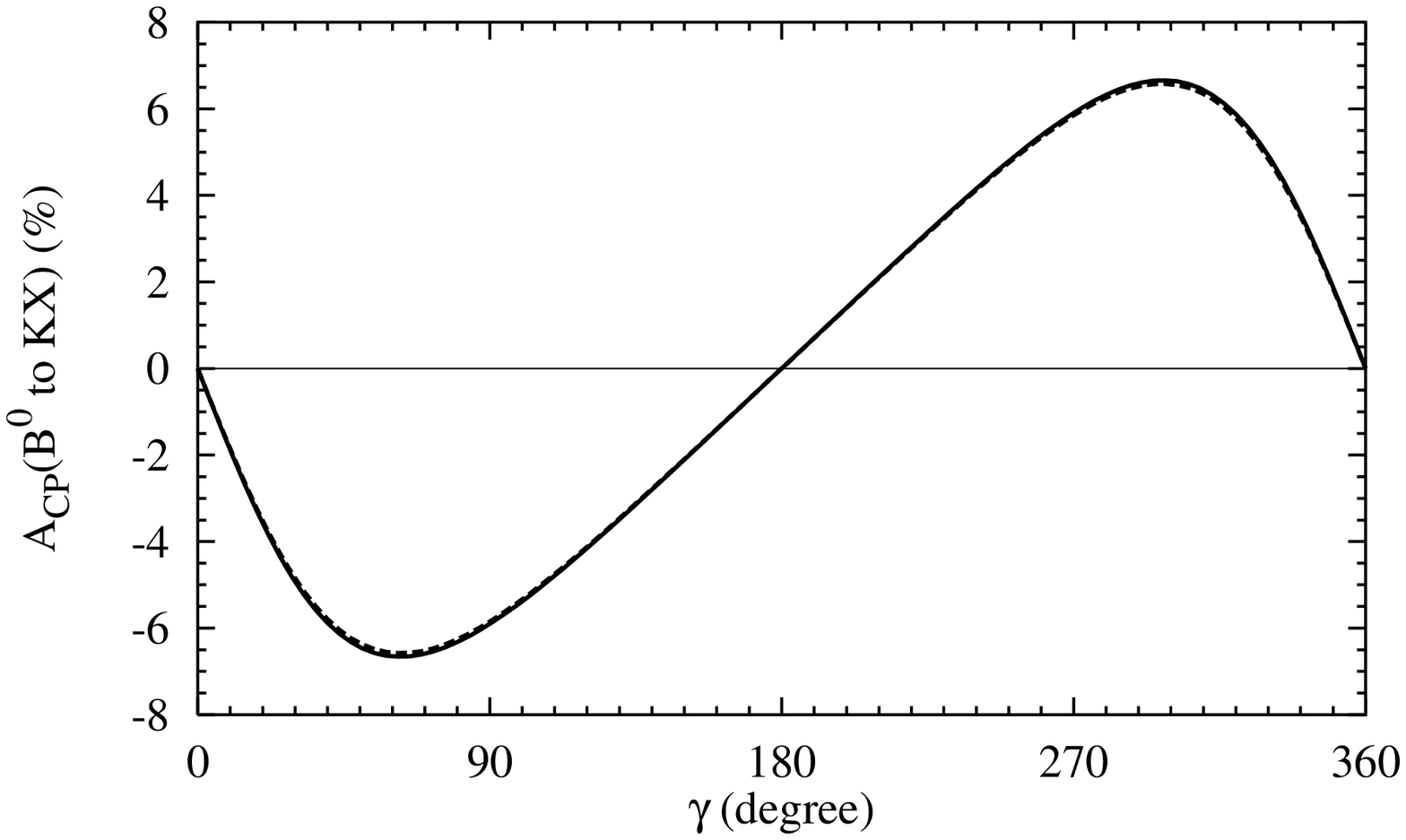}}
\caption {
CP asymmetry in $\bar B^0 \to K^- X$.
} \label{b0cp} 
\end{figure}
\begin{figure}
\centerline{\epsfysize=4truecm \epsfbox{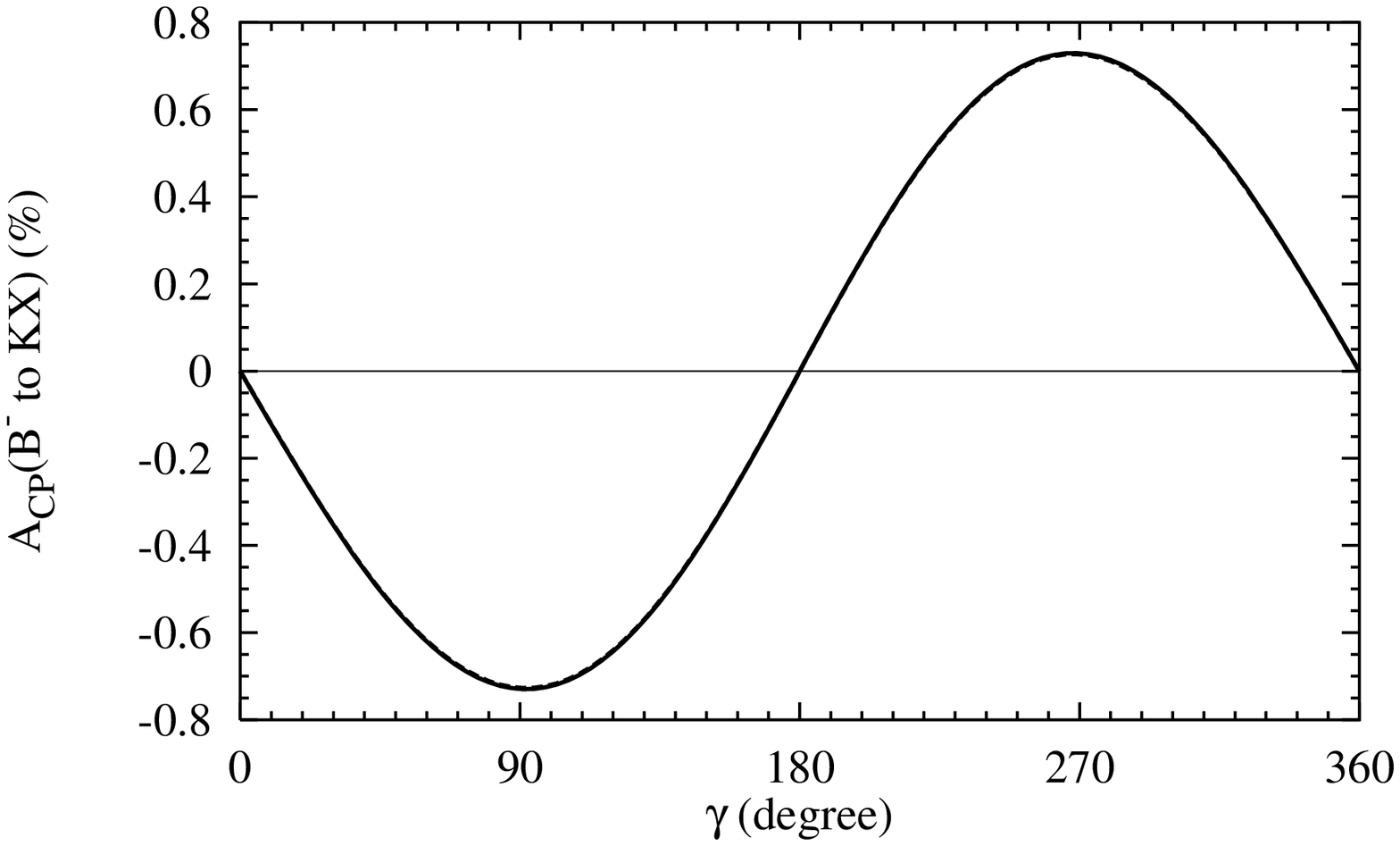}}
\caption {
CP asymmetry in $B^-\to \bar K^0 X$.
} \label{bmcp} 
\end{figure}
\begin{figure}
\centerline{\epsfysize=4truecm \epsfbox{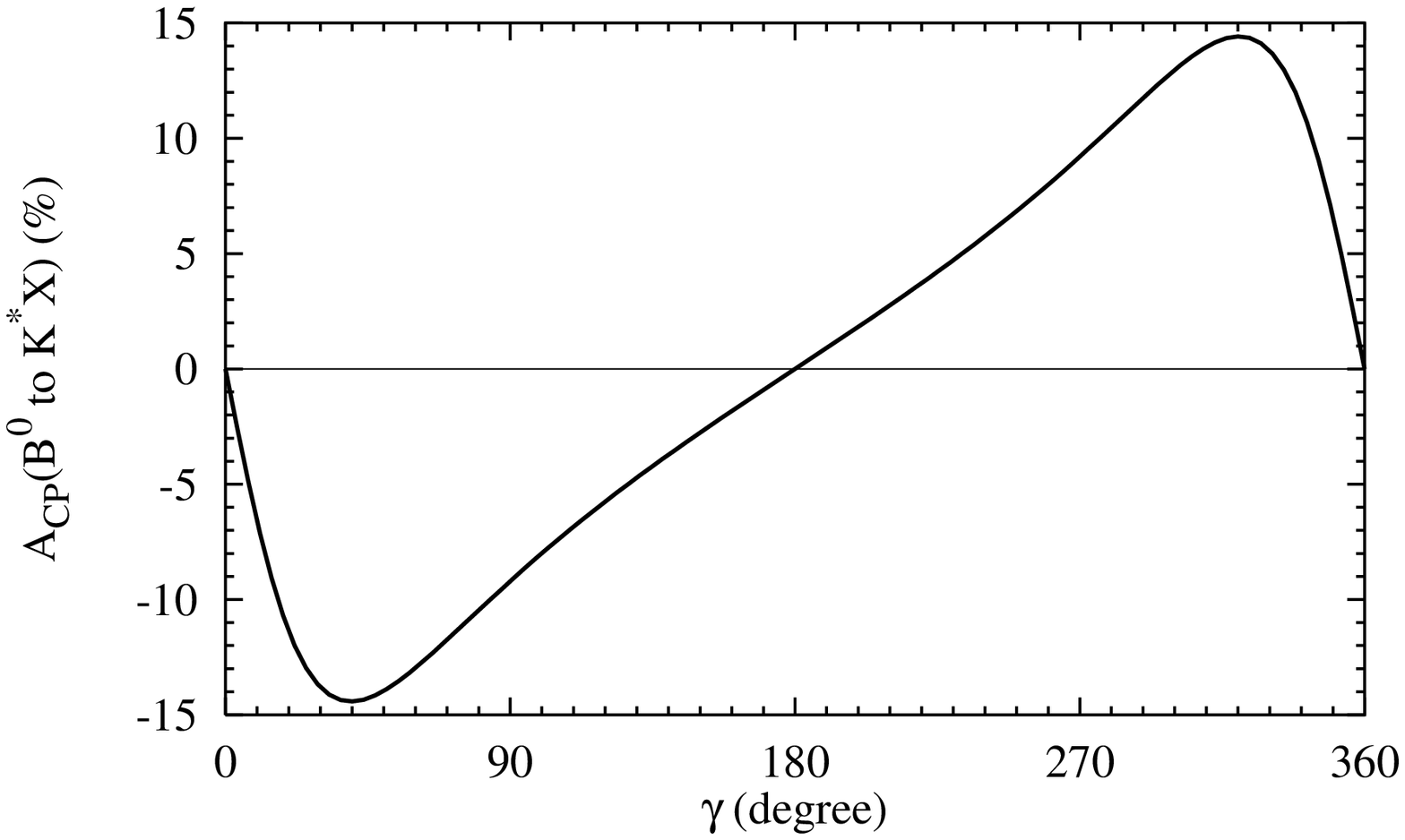}}
\caption {
CP asymmetry in $\bar B^0 \to K^{*-} X$.
} \label{b0starcp} 
\end{figure}
\begin{figure}
\centerline{\epsfysize=4truecm \epsfbox{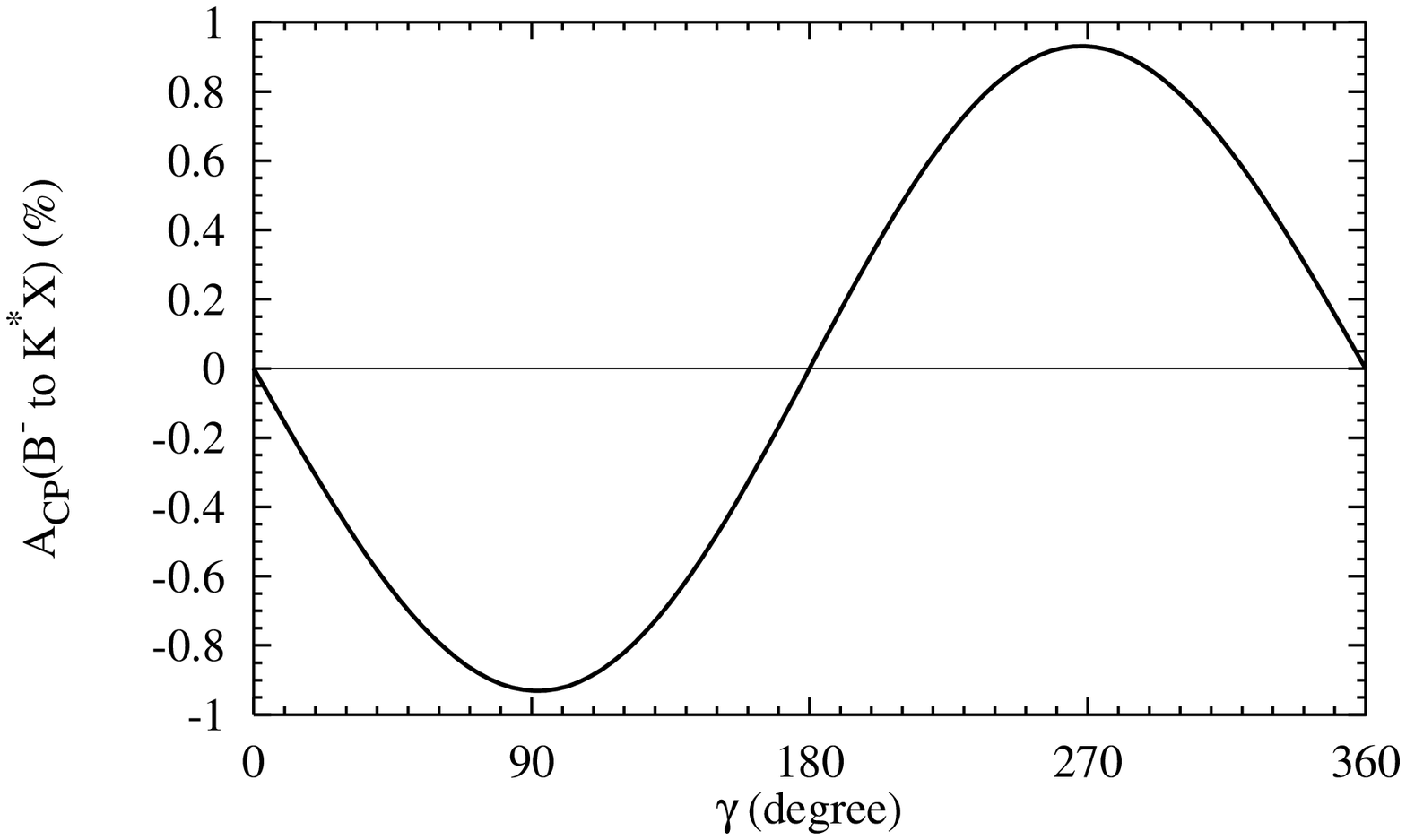}}
\caption {
CP asymmetry in $B^-\to \bar K^{*0} X$.
} \label{bmstarcp} 
\end{figure}


\section*{References}
\begin{thebibliography}{99}
\bibitem{browder}
T.E. Browder, A. Datta, X.-G. He and S. Pakvasa, 
Phys. Rev. {\bf D 57}, 6829 (1998);
D. Atwood and A. Soni, Phys. Rev. Lett. {\bf 81}, 3324 (1998);
Y. Kwon, these proceedings.

\bibitem{bkx} X.-G. He, C.H. Jin and J.P. Ma, hep-ph/0011317.

\bibitem{coef} G. Buchalla, A. Buras and M. Lautenbacher,
Rev. Mod. Phys. {\bf 68}, 1125 (1996);
A. Buras, M. Jamin and M. Lautenbacher, Nucl. Phys. {\bf B 400}, 75 (1993);
M. Ciuchini et al., Nucl. Phys. {\bf B 415}, 403 (1994);
N. Deshpande and X.-G. He, Phys. Lett. {\bf B 336}, 471 (1994).

\bibitem{beneke} M. Beneke, G. Buchalla, M. Neubert and C. Sachrajda, 
Phys. Rev. Lett. {\bf 83}, 1914 (1999);
Nucl. Phys. {\bf B 591}, 313 (2000); hep-ph/0007256.

\bibitem{li} H.-n. Li, these proceedings.

\bibitem{brodsky} S. Brodsky, these proceedings.
 
\bibitem{PDG} Particle Data Group, D.E. Groom et al.,
Eur. Phys. J. {\bf C 15}, 1 (2000).

\bibitem{jp} C.H. Jin and E.A. Paschos, in {\em Proceedings of the
International
Symposium on Heavy Flavor and Electroweak Theory}, Beijing, China,
1995, edited by C.H. Chang and C.S. Huang (World Scientific, Singapore,
1996), p.~132; hep-ph/9504375;
C.H. Jin, Phys. Rev. {\bf D 56}, 2928 (1997);
C.H. Jin and E.A. Paschos, Eur. Phys. J. {\bf C 1}, 523 (1998);
C.H. Jin, Phys. Rev. {\bf D 56}, 7267 (1997);
Phys. Rev. {\bf D 57}, 6851 (1998);
Phys. Rev. {\bf D 62}, 014020 (2000).

\bibitem{gamma} C.H. Jin, Eur. Phys. J. {\bf C 11}, 335 (1999).

\bibitem{hqet} For a review, see M. Neubert, Phys. Rep. {\bf 245}, 259 (1994);
Int. J. Mod. Phys. {\bf A 11}, 4173 (1996).

\bibitem{ma} J.P. Ma, Phys. Lett. {\bf B 488}, 55 (2000).

\end{thebibliography}
\end{document}